\documentclass[12pt]{article}
\title{Hans Bethe and the Global Energy Problems.}
\author{Boris Ioffe\\
Institute of Theoretical and Experimental Physics, \\{\it
\normalsize B.Cheremushkinskaya 25, 117218 Moscow,Russia}}
\date{}
\begin{document}
\maketitle

\newcommand{\be}{\begin{equation}}
\newcommand{\ee}{\end{equation}}

\def\la{\mathrel{\mathpalette\fun <}}
\def\ga{\mathrel{\mathpalette\fun >}}
\def\fun#1#2{\lower3.6pt\vbox{\baselineskip0pt\lineskip.9pt
\ialign{$\mathsurround=0pt#1\hfil##\hfil$\crcr#2\crcr\sim\crcr}}}

\vspace{7mm}

\begin{abstract}

Bethe's view-point on the global energy problems is presented.
Bethe claimed that the nuclear power is a necessity in future.
Nuclear energetic must be based on breeder reactors. Bethe
considered the non-proliferation of nuclear weapons as the main
problem of long-range future of nuclear energetics. The solution
of this problem he saw in heavy water moderated thermal breeders,
using uranium-233, uranium-238 and thorium as a fuel.

\end{abstract}

\bigskip

\section{My contacts with Bethe.}

I heard Bethe's name first time in 1947. I was the student of the
3rd course of the physical faculty of Moscow University. I was not
satisfyed by the level of education there, especially by the
teaching of theoretical physics -- all the best russian
theoreticians -- Landau, Tamm, Leontovich and others were expelled
from the University, because they did not share the official
points of view of Marxist philosophy. After some doubts -- if I am
able to be a theoretical physicist or not -- I decided to try to
give up to Landau his theoretical minimum. I passed successfully
the first entering exam on mathematics and Landau gave me the
program of the whole minimum. That time among now well known
Landau Course of Theoretical Physics only $3\frac{1}{2}$ books
were published: Mechanics, Classical Field Theory, Mechanics of
Condensed Matter and the fist part of Statistics (Classical
Statistics). All other  parts of the Course the student should
study mostly by reading original papers, which were in German and
English. (It was implicitly assumed that the student fluently
knows both languages -- a very nontrivial case at that time.) In
the program of Quantum Mechanics there were 3 papers of H.Bethe:
the theory of atomic levels in the fields of crystall \cite{1};
the theory of collisions of fast electrons  with atoms \cite{2};
the theory of deuteron \cite{3}. Especially, it was very hard to
study the first and the second papers -- about 75 pages each in
German, the language, which I learned only being a boy and almost
forget  and with enormous amount of complicated calculations. I
repeated all of them! Just starting from this moment  I have a
great respect to Bethe.

When I passed all Landau examinations, became a member of Landau
seminar and met Landau very often, as a rule 2--3 times a week, I
realized that he also has a great respect to Bethe. He called him
every time -- Hans Albrecht, not by the last name (Pomeranchuk
called Bethe in the same way.) The comparison with Bethe results
was the highest estimation by Landau.

I would like to tell also about the indirect connection of our
work with one Bethe remark, now almost forgotten. In 1951 the
Pomeranchuk group  at ITEP, in which I participated, started to
work on the project of hydrogen bomb, called ``Tube'' (Truba).
This was the continuation of the work, performed  earlier by
Zeldovich and Landau groups. In US a similar project was developed
by Teller and was known as ``Classical Super''. In short the idea
of the project was the following. A long cylinder is filled by
liquid deuterium. At one end of the cylinder the atomic bomb and
the intermediate  device filled by deuterium and tritium are
situated. It was expected, that after atomic bomb explosion the
fusion reaction will start in $D+T$ mixture, resulting in high
temperatures and the shock wave will propagate along the cylinder,
inducing $D+D$ fusion with the explosion (in principle) of
unlimited power for a long enough tube. The essential step in the
project was the calculation of energy balance. The main source of
energy loss was the bremsstrahlung -- production of $\gamma$-rays
in electron -- ion collisions, since $\gamma$-rays leave the
system. In course of propagation in the cylinder, $\gamma$-rays
undergo Compton scattering. Because the bremsstrahlung spectrum is
softer, than the electron spectrum, the energy of $\gamma$'s
increases at Compton scattering. (Sometime this process is called
inverse Compton scattering.) The calculation of this increasing of
energy loss due to Compton scattering and the calculation of the
energy balance with account of this effect was the main task of
the Pomeranchuk group with collaboration with the Zeldovich group.
Much later, I read in \cite{4}, that at the conference at Los
Alamos in April 1946, where Teller presented the results of his
calculations of Classical Super, Bethe made a remark, that the
account of inverse Compton scattering of $\gamma$'s (not accounted
by Teller), will result in negative energy balance and that the
hydrogen bomb of this type will not explode. In our calculations
after hard work we came to the same conclusion -- the energy
balance was negative. Bethe intuitive  prediction was marvelous!
(This story is described in more details in \cite{5},\cite{6}.)
E.Teller and his group came to the similar conclusion.

I met Hans the first time in 1994, when he visited ITEP. He was
invited  into ITEP Director office, where were also few ITEP
physicists, including myself. According to the standard procedure,
ITEP Director Prof.I.V.Chuvilo told  to the guest about the main
investigations proceeding in the Institute, and suggested the
program of his visit: to see first this installation, then that
and so on. But Bethe refused to follow this program -- he said: I
would like to have a conversation with ITEP theoreticians  and
first of all I would like to know, what they are doing in QCD. So,
we went to my office and for about two hours, I explained Hans our
results in QCD: the calculations of baryon masses, based on the
leading role of spontaneous violation of chiral symmetry in QCD
vacuum, the calculations of baryon magnetic moments etc. Hans
listened  very attentively, put  questions and it was clear, that
he would like to understand the subject completely. Only one time,
after about 1 hour of conversation he said: I would like to rest
for 10 minutes. I left the room, came back in 10 minutes and the
conversation  was continuing. (Hans was 87 at that time!)

\section{The necessity of nuclear power.}

Bethe  interest to nuclear power arises naturally, since starting
from 30-ies he became one of the best specialists in nuclear
physics. It is enough to recall his reviews on theoretical nuclear
dynamics \cite{7}, his paper on diffraction scattering of nucleons
on nuclei \cite{8} his lectures on nuclear theory \cite{9}. In
1942-1946 Bethe was the Head of Theoretical Division at Los
Alamos. In 1956-1957 Bethe participated in experimental work,
where the inelastic cross sections of fission spectrum neutrons
were measured  on various elements \cite{10}. These data are
important for construction of power nuclear reactors, especially
of the breeders on fast neutrons.

Starting from 70-ies Bethe was greatly concerned over global
energy problems -- over the balance of energy production and
consumption in the world in future. In two papers
\cite{11},\cite{12} published in 1976,1977 he discussed these
problems. Bethe expected that the consumption of energy will
steadily increase in next decades, but the production of oil, the
main source of energy, will drop around the year 2000  and the oil
(and gas) prices will jump up. Now we know, that his prediction
about rising of oil and gas  prices was correct and also, what he
did not discussed, few developing countries, like China, India,
Brasil are joining now the club of main energy consumers. Bethe
considered two ways to avoid the future energy crises.

The first one was the energy conservation -- the improving the
efficiency with which the energy is  consumed. He stressed the
necessity  of great efforts towards energy conservation: besides
the technical and industrial progress, public education and tax or
other incentives are needed \cite{10}. However, in this way only a
temporary solution (for the next 10-20 years) of the problem may
be achieved: in the long range perspective the energy conservation
will not help very much.

The second way is the intensive use of new sources of energy.
Among these coal, nuclear fusion and solar power were considered.
Bethe was sceptical about the extensive use of coal, especially in
Europe, where the coal resources are restricted and deep mining is
needed. The use of nuclear fusion requires the solution of many
hard engineering problems. So, Bethe expected \cite{11}, that, in
the best case, fusion might contribute only  few percent of USA's
power supply not to early as  in 2020. The solar power was
considered in more details in \cite{10} and the conclusion was:
solar power is likely to remain extremely expensive in USA, Europe
and Japan, but ``it might be just the right thing for certain
developing countries, which have large desert areas and rather
modest total energy needs''. Bethe  concludes, that all
alternative sources of energy, including wind, biological sources,
should be investigated and developed, but they cannot solve the
future energy problems.

The final conclusion was: there is no alternative to nuclear
power. ``The nuclear power is a \underline{necessity} not merely
an option. A necessity  if we want to make a smooth transition
from our present oil-- and gas--based economies to the post--oil
world'' \cite{12}.

\section{The advantages and problems of nuclear energetics. }

Bethe emphasized that the main problem of long-range future of
nuclear energetics based on  breeder reactors is the problem, how
to provide  the non-proliferation of nuclear  weapons. He shared
the point \cite{13}, that this goal cannot be achieved, if the
facilities, where plutonium or weapon-grade uranium are  produced,
like reprocessing  or isotope  separation plants, would be widely
distributed  over the world. Bethe  believed, that the wide spread
of nuclear energetics may be reconciled  with non-proliferation of
nuclear weapons, if  the fuel for nuclear reactors will be not
only uranium, but also thorium. He supported the proposal by
T.B.Taylor and H.A.Feiverson \cite{13} as a possible scheme of
future nuclear energetics. According to it the nuclear power
plants should be of two types. The first type are the power plants
with thermal reactors using $^{233}U$, $^{238}U$ and thorium as a
fuel. The  moderator in such reactors should be heavy water.  As
the most perspective heavy water reactors. Bethe considered
Canadian reactors CANDU\footnote{In 1995 I participated at
E.Wigner Memorial Meeting of APS, where Bethe presented his
recollections on Wigner. There was also a talk ``Wigner as nuclear
engineer'' given by Alvin Weinberg. After this talk I give a
remark about the development  of nuclear reactors theory in the
USSR (stressing that  it was independent from those un US) and
about construction of heavy water reactors in USSR. After the
Meeting I had a conversation with Bethe -- he did not know about
heavy water reactors in the USSR. (The description of the
development of heavy water reactors in USSR is given in
\cite{14}). } Then one may expect a high, close to 1 conversion
coefficient (the ratio of new formed fission elements -- $^{233}U$
and $^{239}Pu, ~^{241}Pu$ to the burned ones). The ratio of
$^{233}U$ to $^{238}U$ must be chosen in such a way, that after
chemical separation of uranium, it will not be suitable for
nuclear weapons. Also, the concentration of plutonium should be
much smaller than those of $^{233}U$. The nuclear power plants of
the second type are the plants with breeder  reactors on $^{238}U$
and plutonium. The blankets of breeders would be made of thorium
so as  to produce $^{233}U$ for thermal reactors. The  plutonium
produced in thermal reactors would be fed back to breeder
reactors. The number of fast breeders should be much smaller, than
the number of thermal reactors.

The problem of nonproliferation  of nuclear weapons is weaker in
the proposed by Bethe two-component scheme of nuclear energy
plants, than in case of energy production by fast breeder
reactors. The number of breeder  reactors in two-components scheme
is relatively small, they should be under international management
and located in countries of the great internal stability. The
plants for reprocessing breeder blanket should be located in close
proximity to them. Only the thermal reactors would be sold in
international trade. Of course, as mentioned by Bethe, to in order
prevent proliferation, some international agreement is necessary
and, therefore, some restrictions to national  sovereignty. But
this is the price for the world peace.

Bethe expected that nuclear energetics formed in this way will
cover all energy consumption of  the world for many thousand years
and electric energy, produced on nuclear power stations will be
cheaper, than those produced on the stations using fossil fuel.
(It must be mentioned that the last conclusion was based on the
economical calculations performed before Three Mile Island and
Chernobyl accidents. After that the requirements  to the safety on
nuclear power stations strenghtened, essentially in USA, what
resulted in three times increasing of their costs in USA. At the
same time the price of fossil fuel also increased. So, new
economical analysis  must be used now.) In carrying out this
program the problems of chemical separation of the fuel containing
thorium, uranium and plutonium as well as engineering problems
must be solved.  One of the most hard technical problems is the
problem of remote handling of the fuel. The reason is that in the
fuel, containing $^{233}U$ in the reactor core, due to $(n,2n)$
reaction by fast neutrons $^{232}U$ is formed and the products of
radiative chain, started from $^{232}U$, are highly radioactive.
But, on the other side, the same circumstance  provides the
protection against theft.

The other serious problems arising from wide use of nuclear
energetics are: the problem of waste disposal, and the release of
radioactivity by nuclear power stations, especially in case of
fatal accidents on nuclear reactors.

In discussion of the problems of wast disposal Bethe refers to the
Report to the American Physical Society by the Study Group on
Nuclear Fuel Cycles and Waste Management (APS Study Group)
\cite{15}. He cites the general conclusion of the Report:
``Effective long-term isolation for spent fuel, high-level or
transuranic  wastes \underline{can} be achieved by geological
emplacement ... Many waste repository sites with satisfactory
hydrogeology can be identified in continental USA in a variety of
geology formations. Bedded salt can be a satisfactory medium for a
repository , but certain other rock types,  notably granite and
possibly shale, could offer even greater long-term advantages.
Irrespective of the time scale, adopted for reprocessing,
\underline{two} geological demonstration  facilities in
\underline{different} media should be completed''. (Emphasis by
Bethe). The wastes must be solidified, then the solid wastes
should be fused with borosilicate glass and put in stell
cylindres. It is hard to imagine  how the radioactive material
could get out into environment  from the repository at the depth
of 500 m below the ground after such treatment or how the
terrorists could penetrate there.

In some other countries (France, Russia) the decisions  about
waste disposal are close to recommendations, presented in the
Report, although in USA the final decision about waste depository
is not taken till now. It must be mentioned, that in recent years
the proposal of transmutation  by accelerators of radioactive
transuranium elements into nonradioactive ones  is under
discussion and corresponding experiments are under way. Bethe was
strongly interested in this proposal and supported the
construction of accelerator for this purpose.

In discussion of the  problem of radioactivity release by nuclear
power plant, Bethe remarks first, that in case of routine
operation the radioactive  exposure of a person living permanently
near the fence of the power plant is negligable  in comparison
with natural radioactivity. The main danger arises in cases of
serious accidents on reactors. The possibility and the
consequences of such accidents were studied  in 1975 USA Atomic
Energy Commision Reactor Safety Study (the results were presented
in the so called Rasmussen report) and by American Physical
Society's Study Group on Light Water Reactor Safety. Later, in
1986, after the Chernobyl accident a special American panel was
formed for its study. Hans Bethe participated in the APS Study
Group and in the panel.

The Rasmussen report considered the nuclear reactor core meltdown
accident followed by release of radioactivity in the athmosphere,
when the containment failed to keep it inside. The report
estimated the probability of such an event as 1/200000 per 1000
megawatts  electrical reactor year. Of course, this estimation was
purely theoretical, because no such accident was happend before.
The number of fatalities from radiation was estimated as 300. The
APS study did not analyse the probability of such accident, but by
accounting the delayed cancers, came to much higher estimation of
fatalities -- about 10000. Even if we take the APS  estimation,
remarks Berthe \cite{16}, the number of fatalities from the
accident -- about 5 per year for the whole USA -- is not very high
in comparison with other types of non-nuclear accidents.
Naturally, Bethe shared the important conclusions of APS Study
Group \cite{17}: 1) the necessity of improval of  containment
design; 2) mitigation of accident consequences.\\ The main points
of the report of U.S. panel, studyed the Chernobyl accident was
presented by Bethe \cite{18}. Bethe stressed, that
``Chernobyl--type reactor design is seriously flawed''. The
reactor is inherently unstable as a physical system: the power of
reactor increases with increasing of temperature or the content of
steam in the cooling water. This is the main source of this
catastrophic accident. (I would like to mention, that this point
of view completely concides with mine \cite{5,6}). The Chernobyl
reactor  did not have a containment. Therefore,the blow-up of the
reactor  immediately results in crash of the roof of reactor
building and radioactivity  was ejected in the athmosphere. Both
these defects  are absent in U.S. light-water reactors. Moreover,
the containments in U.S. reactors are equipped with water sprays,
which cool down the steam and radioactive products,  condensing
them within the containment. The most dangerous fission products
-- iodine and cesium (in the form of cesium iodide) are highly
soluble in the water, ejected by sprays. For these reasons,
remarks Bethe, in Three Mile Island accident it was released to
the enviroment about one-millionth of the amount of iodine that
was ejected into athmosphere at Chernobyl. No amounts of cesium or
strontium were detected at Three Mile Island accident. Bethe
concludes, that the likelihood of seriously damaging accident is
clearly much smaller for  U.S. light-water reactors than for the
Chernobyl-type. But both the accident at Chernobyl and the less
serious one at Three Mile Island indicate ``that continued concern
for safety is maintained over the full lifetime of U.S. plants''
\cite{17} (I add: and over the whole world).

I am thankful to G.Brown for information about Bethe's activity in
nuclear  energetics and to U.Meissner for his hospitality at the
Bonn University, where this paper was written.

This work was  supported in part by U.S. CRDF Cooperative Grant
Program, Project RUP2-2621-MO-04, RFBR grant 03-02-16209 and the
funds from EC to the project ``Study of Strongly Interacting
Matter'' under contract 2004 No R113-CT-2004-506078.

\end{document}